\documentclass{article}
\usepackage[utf8]{inputenc}
\usepackage[table]{xcolor}
\usepackage{physics}
\usepackage{amsmath}

\usepackage{url}
\usepackage[hidelinks]{hyperref}
\hypersetup{breaklinks=true}
\usepackage[
backend=biber,
sorting=none
]{biblatex}

\title{Social Advantage with Mixed Entangled States}
\author{Aritra Das\footnote{Indian Institute of Technology Kanpur, Uttar Pradesh, India} \, and Pratyusha Chowdhury\footnote{St. Xavier's College, Kolkata, West Bengal, India}}
\date{\vspace{-5ex}}

\begin{document}

\maketitle

\begin{abstract}
It has been extensively shown in past literature that Bayesian Game Theory and Quantum Non-locality have strong ties between them. Pure Entangled States have been used, in both common and conflict interest games, to gain advantageous payoffs, both at the individual and social level. In this paper we construct a game for a Mixed Entangled State such that this state gives higher payoffs than classically possible, both at the individual level and the social level. Also, we use the I-3322 inequality so that states that aren't helpful as advice for Bell-CHSH inequality can also be used. Finally, the measurement setting we use is a Restricted Social Welfare Strategy (given this particular state).
\end{abstract}

\section{Introduction}

The quantum theory emerged when most of the physicists realized that physics at the atomic level could not be completely described by classical mechanics. Plank was the first to give the notion of "quanta", which was further developed by Einstein. Though Heisenberg and Bohr, the further creators of the theory, believed in the innate uncertainty of the behaviour of atoms, Einstein never accepted it. Einstein was always against the \textit{Copenhagen interpretation} of quantum mechanics. Thus Einstein-Podolsky-Rosen put forward the EPR paradox in their paper in 1935 \cite{epr}, that claimed quantum mechanics is incomplete,that is, it doesn't provide a complete picture of the physical reality.

To resolve this, Einstein - Podolsky - Rosen  introduced a \textit{Hidden Variable Theory}, in their paper , which according to them removed all the in-determinism and accounted for the behaviour of the observables in quantum mechanics. This led Bohr to publish a paper in the same journal under the same name, where he stated that the criterion of physical reality given by EPR contains an essential ambiguity when applied to quantum phenomenon \cite{Bohr}. Hence, there continued a debate between Einstein and Bohr regarding the fundamental nature of reality.

But later, Bell formulated an inequality \cite{Bell} which was satisfied by all local realistic theories. Eventually, the quantum violation of Bell’s inequality proved that no \textit{Local Realistic Hidden Variable Theory} can exist from which quantum mechanics can be derived. The value by which quantum mechanics violates a particular Bell Inequality
is called the \textit{Tsirelson bound} for that particular Bell Inequality. The Tsirelson bound for Bell CHSH \cite{CHSH} is 2$\sqrt{2}$. 

Quantum states that violate Bell's inequality are all non-local states (entangled). However this leads to the question:

\begin{quote} Is Bell's Inequality sufficient to show non-locality (or entanglement) ?
\end{quote}

It turns out that Bell's inequality is not sufficient to prove non-locality. The states that violate Bell's inequality are definitely non-local, but there are other states that do not violate a particular bell inequality but are still non-local. So in such cases, other inequalities are required which are inequivalent to this inequality. In this paper, we have worked with one such inequality - the I3322 inequality \cite{3322}.

\section{Game Theory}
Game theory is mathematical modeling of strategic interaction among rational beings, used widely in economics \cite{book1} , political sciences \cite{book2} , biological phenomena \cite{book3}, as well as logic, computer science and psychology \cite{book4}. It is the study of human conflict and cooperation, or in other words the study of optimal decision making of different players, each with a set of action having particular payoffs.It is the payoff which decides the preference of an action over another. Von Neumann and Morgenstern \cite{book5} were the pioneers of game theory.    

Game theory can be cooperative (common interest) or non-cooperative (conflict interest). Cooperative game includes competition between groups whereas non-cooperative game includes analyzing strategies and payoffs of individual players using the concept of \textit{Nash Equilibrium }\cite{nash}. In a game if a player chooses a unique action from a set of  available action it is called pure strategy,but if a probability distribution over a set of action is available it is called mixed strategy.Nash proved that in any game with finite number of action for each player there is always a mixed strategy Nash Equilibrium. Later the concept of \textit{Bayesian Games} i.e, games of Incomplete Information was introduced \cite{harsanyi} and Aumann proved the existance of correlated equilibria \cite{aumann} in these games, as opposed to Nash Equilibria. 

\section{Quantum Game Theory}
\subsection{Non-locality and Bayesian Game Theory}
Non-locality is one of the most counter intuitive aspects of Quantum Mechanics. The principle of locality states that an object can only be affected by it's immediate surroundings and not by remote or distant objects. However Quantum theory is not consistent with this and is inherently non-local in nature, unlike the rest of classical physics. For example, two entangled particles placed far apart, can display correlations inexplicable by classical physics. These correlations can not be as a result of a signal transfer as that would imply superluminal communication. On the other hand, in 1964, Bell \cite{Bell} showed that these correlations can not be as a result of pre-decided strategies either.

In 2013, Brunner and Linden \cite{Brunner} demonstrated strong links between Quantum Nonlocality and Bayesian Game Theory. Specifically they showed that the normal form of a Bayesian game is equivalent to a Bell Inequality test scenario. They showed that when the two players in the game share non local resources such as entangled pair of quantum particles, they can outperform players using any sort of classical resources. This can happen, for example, when the payoff function of the players corresponds to a Bell Inequality, like the CHSH inequality \cite{CHSH}, as first discussed by Cheon and Iqbal \cite{Cheon} but also when the payoff function doesn't correspond to any Bell Inequality. They showed that more generally, for Bayesian Games, Quantum Mechanics provides clear and indisputable advantage over all classical resources.

\subsection{Nonlocality in Conflict Interest Games}
 Brunner and Linden showed that Quantum Mechanics indeed provides an advantage over classical resources for all Bayesian Games, but the examples they provided were all Common Interest Games (games where it is beneficial for both the players to cooperate rather than oppose each other). In fact, until 2015, all other known non local games, including the GHZ-Mermin game \cite{GHZ}, the Bell-CHSH game \cite{CHSH}, and the Hidden Matching game \cite{Mermin, Peres} were all examples of Common Interest Games  (mostly, because the average payoff functions for both Alice and Bob were the same).
 
 In 2015, however, Anna Pappa et al. \cite{Anna} demonstrated that quantum advice can offer an advantage compared to classical advice even in conflicting interest games. They explicitly constructed an incomplete information game with conflicting interests, where quantum strategies yielded fair equilibria with average payoffs strictly higher than those achievable by classical means, for both the players. 
 
\subsection{Fair and Unfair Strategies}
Classical equilibria can be of two types - 
\begin{itemize}
    \item Fair equilibria, where the average payoffs for both the  players are equal, and 
    \item Unfair equilbria, where the payoffs for the players are unequal.  
\end{itemize}
Up until 2016, most of the games (both Common Interest and Conflicting Interest) proposed, dealt with Fair equilibria - that is, they showed that quantum fair payoffs surpass classical fair equilibrium payoffs. In 2016, Arup Roy, Amit Mukherjee et al. \cite{ISI1} showed that quantum strategies can outperform not only fair classical equilibrium strategies but unfair strategies too. They analytically characterized some non-local correlations, that would yield unfair average payoffs strictly higher than the classical ones in Anna Pappa's game.
\subsection{Social Welfare Solutions and Pure Entangled States}
Until now, we've been talking about equilibria for individual players - that is, states where the players can't increase their payoffs further, by unilaterally changing their individual strategies. Such equilibria are called correlated equilibria (as opposed to Nash Equilibria). Psychological factors indicate that sometimes, instead of focusing solely on their individual payoffs, players may also consider additional social goals - one such idea is the Social Welfare Solution (SWS).

In such a strategy, players aim to maximize the sum of their individual payoffs. Out of all the possible quantum strategies, the ones that increase the sum of the payoffs (above the classical value) are called Quantum Social Welfare Solution (Quantum-SWS) and the quantum state producing this strategy is called Quantum Social Welfare Advice (Quantum-SWA). 

In 2017, Manik Banik, Some Sankar Bhattacharya et al. \cite{ISI2} showed that any two-qubit pure entangled state can act as Quantum-SWA for some Bayesian Game. Hence given any pure entangled state between two qubits, there exists at least one game where this state provides Quantum-SWS.

\section{Mixed Entangled States and I3322 Inequality}
The above discussion raises the question \begin{quote} Can Mixed Entangled States can be used as Quantum-SWA for Bayesian Games?\end{quote} 
We answer this question in the affirmative, by explicitly constructing a Bayesian game, where a mixed entangled state gives higher unfair payoffs and higher social payoffs than classical values.

\subsection{The Premise}
The setup is now changed - it is no more a 2-2-2 scenario. Now, there are 2 players with 3 possible questions (3 possible measurement settings) with 2 actions (or outcomes) to each question $\in\{0,1\}$.
\newline

Let the players be $A$ for Alice and $B$ for Bob. They can be asked 3 questions, each namely: $\{A_1, A_2, A_3\}$ and $\{B_1, B_2, B_3\}$. Their answers (or outcomes) for these are $\in\{0,1\}$. So, $P(x,y\vert A_i, B_j)$ refers to the probability that, when $A$ is asked $A_i$ and $B$ is asked $B_j$ they reply $x$ and $y$ as answers, respectively, with $x,y \in \{0,1\}$.

First, we construct the classical game.

\subsection{The Classical Game}
\subsubsection{Classical Strategies}
A classical strategy means $A$ and $B$ both locally predecide their answers to the questions. For each question, $A$ or $B$ can answer either $0$ or $1$. So for the entire set of $3$ questions, there can be $2^3=8$ different set of answers. Each such set of answers is called a strategy for that particular player. For example if A answers 0 to all questions (that is $0$ for $A_1$, $0$ for $A_2$ and $0$ for $A_3$) then her answer strategy is $000$.

Strategies are named by considering the answer sequence (for $A_1, A_2, A_3$ or $B_1, B_2, B_3$, in this order) as a binary string and converting it into decimal. So $000$ becomes $g_0$ and $010$ becomes $g_2$ and so on. So the strategies for each individual player ($A$ or $B$) are $g_0, g_1, \dots g_6, g_7$

\subsubsection{Probability Boxes}

The Probability Box (or local box) is a table of numbers, which shows the how the strategy relates the questions to the answers. It shows for each question, with what probability a player chooses a particular answer. There is a one to one relation between the strategy ordered pair $(g_A, g_B)$ and the probability box.

In this part, we use classical probability boxes, that is local boxes. Local boxes respect locality, that is the probability that $A$ gives a particular answer to some question is independent of what $B$ is asked and what his response is. A general classical probability box looks takes the following form:

\begin{center}
\begin{tabular}{c|cccc}
&	$OO$ &	$O1$ &	$1O$ &	$11$ \\
\hline
$A_1B_1$ &	$C_{11}$ &	$M_1 - C_{11}$ &	$N_1-C_{11}$ &	$1-M_1-N_1+C_{11}$ \\
$A_1B_2$ &	$C_{12}$	& $M_1 - C_{12}$	 & $N_2-C_{12}$ &	$1-M_1-N_2+C_{12}$\\
$A_1B_3$ &	$C_{13}$	& $M_1-C_{13}$	 & $N_3-C_{13}$ &	$1-M_1-N_3+C_{13}$ \\
$A_2B_1$ &	$C_{21}$	& $M_2-C_{21}$	 & $N_1-C_{21}$ &	$1-M_2-N_1+C_{21}$ \\
$A_2B_2$ &	$C_{22}$	& $M_2-C_{22}$	 & $N_2-C_{22}$ &	$1-M_2-N_2+C_{22}$ \\
$A_2B_3$ &	$C_{23}$	& $M_2-C_{23}$	 & $N_3-C_{23}$ &	$1-M_2-N_3+C_{23}$ \\
$A_3B_1$ &	$C_{31}$	& $M_3-C_{31}$	 & $N_1-C_{31}$ &	$1-M_3-N_1+C_{31}$ \\
$A_3B_2$ &	$C_{32}$	& $M_3-C_{32}$	 & $N_2-C_{32}$ & $1-M_3-N_2+C_{32}$ \\
$A_3B_3$ & $C_{33}$	& $M_3-C_{33}$	 & $N_3-C_{33}$ &	$1-M_3-N_3+C_{33}$ \\
\end{tabular}
\end{center}

Row $A_iB_j$ and column $xy$ represents probability of answering $(x,y)$ for the question $A_iB_j$, that is $P(x,y\vert A_iB_j)$. It is easy to see that this is a local box, since for example, the probability that $A$ answers 0 to $A_1B_1$ is $P(00\vert A_1B_1) + P(01\vert A_1B_1) = M_1$ which is the same as the probability that $A$ answers $0$ to $A_1B_2$ or $A_1B_3$, etc.So the probability that $A$ given question $A_1$ answers $0$ is the same independent of what question $B$ is asked.

\subsubsection{Utility Boxes}

For a particular game each player chooses one out of these 8 strategies available to them. While the strategy dictates the move or answer that the player gives upon being asked the question, the reward or payoff he gets from that answer is described by the Utility Box. Given the pair of strategies that are being used, the player’s individual payoffs can be calculated from the Utility Boxes.

A’s answers are listed along the columns and B’s along the rows. For a question pair $(x, y)$, the ordered pair $(u_1,u_2)$ in row $a$ and column $b$  represents the payoff $A$ and $B$ get, respectively, on answering with $a$ and $b$. We designate $A$'s reward as $u_A(a,b\vert x,y) := u_1$ and $B$'s as $u_B(a,b\vert x,y) := u_2$.

For questions $A_1B_1, A_1B_2, A_1B_3$ \begin{tabular}{c|cc}
 & 0 & 1 \\
\hline 
0 & $\frac23, 1$ & $-\frac13,0$ \\
1 & $0,\frac13$  & $0,\frac13$ 
\end{tabular}

For questions $A_2B_1$ and $A_3B_1$ \begin{tabular}{c|cc}
 & 0 & 1 \\
\hline 
0 & $\frac12, 0$ & $\frac12,0$ \\
1 & $-\frac12,-1$  & $\frac12, 0$ 
\end{tabular}

For questions $A_2B_3$ and $A_3B_2$ \begin{tabular}{c|cc}
 & 0 & 1 \\
\hline 
0 & $-\frac23, -\frac13$ & $\frac13, \frac23$ \\
1 & $\frac13, \frac23$  & $\frac13, \frac23$ 
\end{tabular}

For question $A_2B_2$ \begin{tabular}{c|cc}
 & 0 & 1 \\
\hline 
0 & $\frac13, \frac23$ & $\frac13, \frac23$ \\
1 & $-\frac23, -\frac13$  & $\frac13, \frac23$ 
\end{tabular}

For question $A_3B_3$ \begin{tabular}{c|cc}
 & 0 & 1 \\
\hline 
0 & $0, 0$ & $-\frac13,\frac13$ \\
1 & $\frac13,-\frac13$  & $0,0$ 
\end{tabular}

\subsubsection{Classical Payoffs}
For each pair of strategy $(g_A, g_B)$ that Alice and Bob choose, they each get a particular payoff or reward. 

Payoff for a particular question is calculated by using the strategy (Probability Box) to find the player’s answer and then using the utility box to find the corresponding payoff. Average payoff is the average of the individual payoff over all possible questions. 

$$F_A = \sum_{a,b,x,y} p(x,y) P(a,b\vert x,y) u_A(a, b, x, y)$$
$$F_B = \sum_{a,b,x,y} p(x,y) P(a,b\vert x,y) u_B(a, b, x, y)$$

Where, $p(x,y)$ is the probability that the question pair $(x,y)$ is asked. 

In our game, we assume each question is equally likely to be asked, then $p(x,y) = \frac{1}{N^2} \forall (x,y)$ where $N$ is the number of questions. Here, there are 3 questions for each party ($A_1, A_2, A_3$ or $B_1, B_2, B_3$). So, $p(x,y) = \frac{1}{9} \forall x, y$.

The classical payoffs, using the Probability Box for $p(a,b\vert x,y)$ and the Utility Boxes for $u_A$ and $u_B$, are then:

\begin{align*}
F_A = &\frac19 (C_{11} + C_{12} + C_{13} + C_{21} + C_{31} -C_{23} - C_{32} + C_{22} \\ 
&-M_1 -2N_1 - N_2 -\frac{M_3}{3} + \frac{N_3}{3} + 2) \\
\\
F_B = &\frac19 ( C_{11} + C_{12} + C_{13} + C_{21} + C_{31} -C_{23} - C_{32} + C_{22} \\
&-M_1 -2N_1 - N_2 +\frac{M_3}{3} - \frac{N_3}{3}+3 ) \\
\end{align*}

The $C_{ij}, M_i, N_j$ values ($\in\{0,1\}$ for classical strategies) are decided by the strategies $g_i$. Also, since the expressions for $F_A$ and $F_B$ are different, the payoffs are unfair.

\subsubsection{Classical Equilibria}

Since each of the players has a choice of 8 different strategies ($g_0, g_1, \dots g_7$), the final payoff box is a $8\times8$ table, with the first entry being the payoff for Alice and the second one being the payoff for Bob. A factor of $\frac{1}{27}$ has been ignored in the table, to keep things cleaner.

\begin{center}
\begin{tabular}{c|cccccccc}
&	g0	& g1 & g2 & g3 & g4 & g5 & g6 &	g7 \\
\hline
g0 & 6, 9 & 5, 10 & 6, 9 & \cellcolor{yellow!30} 5, 10 & 3, 6 & 2, 7 & 3, 6 & 2, 7 \\
g1 & 7, 8 & \cellcolor{yellow!30} 6, 9 & 4, 5 & 3, 6	& 7, 8 & \cellcolor{yellow!30} 6, 9 & 4, 5 & 3, 6 \\
g2 & 3, 6 & -1, 4 & \cellcolor{yellow!30} 6, 9 & 2, 7 & 3, 6 & -1, 4 & 6,9 & 2,7 \\
g3 & 4, 5 & 0, 3 & 4, 5 & 0, 3 & \cellcolor{yellow!30} 7, 8 & 3, 6 &	\cellcolor{yellow!30} 7, 8 & 3, 6 \\
g4 & 0, 3 & 2, 7 & 3, 6 & \cellcolor{yellow!30}5, 10 & 0, 3 & 2, 7 & 3, 6	& 5, 10 \\
g5 & 1, 2	& 3, 6 & 1, 2 & 3, 6	& 4, 5 & \cellcolor{yellow!30} 6, 9 & 4, 5 & \cellcolor{yellow!30} 6, 9 \\
g6 & -3, 0 & -4, 1 & 3, 6 & 2, 7 & 0, 3 & -1, 4 & 6, 9 & 5, 10 \\
g7 &	-2, -1 & -3, 0 & 1, 2 & 0, 3 & 4, 5 & 3, 6 & 7, 8 & \cellcolor{yellow!30} 6, 9 \\
\end{tabular}
\end{center}

The equilibria (all are biased / unfair) have been shaded yellow. These are the stable states for this game. Also, social welfare solution payoff is $(\frac{15}{27}, \frac{15}{27})$. Our next task is to check whether a quantum strategy can increase payoffs of the individual parties above the classical values. 

\subsection{The Quantum Game}

Now, we devise a means to play this game using a quantum state. In this scenario the two players share a Mixed Entangled State. They are asked questions $A_1, A_2, A_3$ and $B_1, B_2, B_3$ respectively, and they get their answer by performing suitable measurements on the shared state. The objective is to generate a payoff for both players, that exceeds the classical equilibrium payoffs. 

We do this by implanting a quantum inequality in the payoff function so that quantum processes can exceed the upper bound for classical processes and hence produce payoffs higher than all classical payoffs.

\subsubsection{The Inequality}

We choose the I-3322 inequality. This inequality was discovered in 2003 by Daniel Collins and Nicolas Gisin \cite{3322}, but little work was done on it, other than finding it's maximal violation value using infinite dimensional quantum systems in 2010 \cite{maxviol}.

The important thing about this inequality is that it is \textbf{inequivalent} to the Bell-CHSH inequality. This means that there are states that don't violate Bell-CHSH inequality but violate this.

The inequality is usually represented in the following way:

\begin{center}
\begin{tabular}{c|ccc}
& -1 & 0 & 0\\
\hline
-2 & 1 & 1 & 1 \\
-1 & 1 & 1 & -1 \\
0 & 1 & -1 & 0 \\
\end{tabular}
\end{center}

where the numbers correspond to the coefficients of:

\begin{center}
\begin{tabular}{c|ccc}
& $P(A_1)$ & $P(A_2)$ & $P(A_3)$  \\
\hline
$P(B_1)$ & $P(A_1B_1)$ & $P(A_2B_1)$ & $P(A_3B_1)$ \\
$P(B_2)$ & $P(A_1B_2)$ & $P(A_2B_2)$  & $P(A_3B_2)$\\ 
$P(B_3)$ & $P(A_1B_3)$ & $P(A_2B_3)$  & $P(A_3B_3)$\\ 
\end{tabular}
\end{center}

in the inequality. Here, for succinctness, we write $P(00\vert A_iB_j)$ as $P(A_iB_j)$ and $P(0\vert A_i)$ as $P(A_i)$.

Rewriting the inequality in a form closer to that of the CHSH inequality, we get

\begin{equation*}
\begin{split}
S=-\frac{1}{3}P(01\vert A_1B_1)-\frac{2}{3}P(10\vert A_1B_1)+\frac{1}{3}P(00\vert A_1B_2)-\frac{1}{3}P(01\vert A_1B_2) \\
-\frac{1}{3}P(10\vert A_1B_2)+\frac{2}{3}P(00\vert A_1B_3)-\frac{1}{3}P(01\vert A_1B_3)+\frac{1}{3}P(00\vert A_2B_1) \\
-\frac{2}{3}P(10\vert A_2B_1)+\frac{2}{3}P(00\vert A_2B_2)-\frac{1}{3}P(10\vert A_2B_2)-P(00\vert A_2B_3) \\
+\frac{1}{3}P(00\vert A_3B_1)-\frac{2}{3}P(10\vert A_3B_1)-\frac{4}{3}P(00\vert A_3B_2)-\frac{1}{3}P(10\vert A_3B_2)\\
\end{split}
\end{equation*}

And then plugging in the variables from the Probability Box gives

$$S = C_{11}  + C_{12}+ C_{13} + C_{21}+ C_{22} -C_{23} + C_{31}- C_{32}  - M_1 -2N_{1} -N_2$$

\subsubsection{Maximum Violation of the Inequality}

For all classical systems, I-3322 satisfies $S\leq0$. For quantum mechanical systems however, a numerical optimization suggests that the maximum value is 0.25 \cite{3322}. The same is suggested by another approach using infinite dimensional quantum systems  \cite{maxviol}.

The state that produces this maximum value is the Maximally Entangled Bell State $\ket{\Psi^-}$:
$$\ket{\Psi^-} = \frac{1}{\sqrt{2}}\left(\ket{01}-\ket{10}\right)$$

Choosing appropriate measurements for A and B, gives the value of the inequality $S_{\ket{\Psi^-}} = 0.25$.

Also, since $F_A + F_B = \frac{1}{9}\left(2S+5\right)$, and this state gives the maximum possible value of $S$, this state is automatically the Social Welfare Solution for this game.

However, since this inequality is in-equivalent to the Bell-CHSH inequality, there exist states that violate this inequality but not the Bell-CHSH inequality. We choose one such Mixed Entangled State and corresponding measurements, with the aim to increase the payoffs beyond classical limits.

\subsubsection{The Quantum State}

The state shared between the $A$ and $B$ is the following mixed entangled state:

$$\rho_{AB} = 0.85\ket{\Phi}\bra{\Phi}+0.15\ket{01}\bra{01}$$
where 
$$\ket{\Phi} = \frac1{\sqrt{5}} \left(2\ket{00} + \ket{11}\right)$$

The Density Matrix for state $\rho_{AB}$ is:

$$\rho_{AB} = 
\begin{bmatrix}
0.68 & 0 & 0 & 0.34 \\
0 & 0.15 & 0 & 0 \\
0 & 0 & 0 & 0 \\
0.34 & 0 & 0 & 0.17 \\
\end{bmatrix}
$$

\subsubsection{The Measurements}

The 6 questions in the classical game have their corresponding measurements for the quantum version. These are all projective measurements, specified by their polar and azimuthal angles $(\theta, \phi)$. The probabilities for these measurements are calculated by applying the density matrix of the proper eigenvalue of the measurement operator on the density matrix of the quantum state and then taking trace.

\begin{align*}
    A_1 &\equiv (\eta, 0) \\
    A_2 &\equiv (-\eta, 0) \\
    A_3 &\equiv (-\frac{\pi}{2}, 0) \\
    \\
    B_1 &\equiv (-\chi, 0) \\
    B_2 &\equiv (\chi, 0) \\
    B_3 &\equiv (\frac{\pi}{2}, 0) \\
\end{align*}

such that $\cos\eta = \sqrt{\frac{7}{8}}$ and $\cos\chi=\sqrt{\frac{2}{3}}$.

Applying the measurements, with the appropriate eigenvalues, we can find out all the elements of the probability box:
\begin{center}
\begin{tabular}{lll}
$M_1  = 0.808687$, & $M_2=  0.808687$, & $M_3 = 0.5$ \\
$N_1 =  0.646969$, & $N_2 =  0.646969$, & $N_3 = 0.5$ \\
$C_{11} = 0.576785$, & $C_{12} = 0.646188$, & $C_{13} =  0.464447$, \\
$C_{21} = 0.646188$, & $C_{22} = 0.576785$, & $C_{23} = 0.344239$, \\
$C_{31} = 0.421634$, & $C_{32} = 0.225335$, & $C_{33} =  0.08$ \\
\end{tabular}
\end{center}

\subsubsection{Quantum Payoffs}
The quantum payoffs are then calculated using the same formulas as those for classical payoffs:

\begin{align*}
F_A = &\frac19 (C_{11} + C_{12} + C_{13} + C_{21} + C_{31} -C_{23} - C_{32} + C_{22} \\ 
&-M_1 -2N_1 - N_2 -\frac{M_3}{3} + \frac{N_3}{3} + 2) \\
= &\frac{6.03858}{27}\\
\end{align*}
\begin{align*}
F_B = &\frac19 ( C_{11} + C_{12} + C_{13} + C_{21} + C_{31} -C_{23} - C_{32} + C_{22} \\
&-M_1 -2N_1 - N_2 +\frac{M_3}{3} - \frac{N_3}{3}+3 ) \\
= &\frac{9.03858}{27}
\end{align*}

The quantum payoff value $(\frac{6.03858}{27}, \frac{9.03858}{27})$ is greater than the classical equilibrium value $(\frac{6}{27}, \frac{9}{27})$.

Also, the social welfare value $\frac{15.0772}{27}$ exceeds that for all classical equilibria $\left(\frac{15}{27}\right)$.

\section{Conclusion}
We have, hence, constructed a game where a mixed entangled state provides higher individual payoffs than the classical equilibria. The social welfare payoff is also increased beyond the upper limit for the classical scenario. 

Finally, if we restrict the advice that the referee gives to the players as this particular mixed state, the measurement setting chosen maximizes the Social Welfare Value. Hence, it can be thought of as a Restricted Social Welfare Solution.
 
\setcounter{biburllcpenalty}{7000}
\setcounter{biburlucpenalty}{8000}
\printbibliography
 
\end{document}